\documentclass[11pt,onecolumn,english,preprintnumbers,amsmath,amssymb,floatfix,nofootinbib]{revtex4}
\usepackage{tikz,xcolor}
\usepackage[colorlinks = true,  
linkcolor = blue,
urlcolor  = blue,
citecolor = blue,
anchorcolor = blue]{hyperref} 

\definecolor{lime}{HTML}{A6CE39}
\DeclareRobustCommand{\orcidicon}{%
	\begin{tikzpicture}
	\draw[lime, fill=lime] (0,0) 
	circle [radius=0.16] 
	node[white] {{\fontfamily{qag}\selectfont \tiny ID}};
	\draw[white, fill=white] (-0.0625,0.095) 
	circle [radius=0.007];
	\end{tikzpicture}
	\hspace{-2mm}
}

\foreach \x in {A, ..., Z}{%
\expandafter\xdef\csname orcid\x\endcsname{\noexpand\href{https://orcid.org/\csname orcidauthor\x\endcsname}{\noexpand\orcidicon}}
}

\usepackage[T1]{fontenc}
\usepackage[latin9]{inputenc}
\usepackage{color}
\usepackage{array}
\usepackage{amstext}
\usepackage{graphicx}
\usepackage{esint}
\usepackage{rotating}
\usepackage{appendix}
\usepackage{float}
\usepackage{subcaption}

\usepackage[font=small,labelfont=bf]{caption}
\usepackage{xcolor}
\usepackage{ulem}

\makeatletter

\@ifundefined{textcolor}{}
{%
 \definecolor{BLACK}{gray}{0}
 \definecolor{WHITE}{gray}{1}
 \definecolor{RED}{rgb}{1,0,0}
 \definecolor{GREEN}{rgb}{0,1,0}
 \definecolor{BLUE}{rgb}{0,0,1}
 \definecolor{CYAN}{cmyk}{1,0,0,0}
 \definecolor{MAGENTA}{cmyk}{0,1,0,0}
 \definecolor{YELLOW}{cmyk}{0,0,1,0}
 }

\@ifundefined{definecolor}
 {\usepackage{color}}{}
\@ifundefined{definecolor}
 {\usepackage{color}}{}
\makeatother
\usepackage{babel}

\newcommand{\beq}{\begin{equation}}
	\newcommand{\eeq}{\end{equation}}

\newcommand{\cL}{\mathcal{L}}

\def\Re{{\cal R \mskip-4mu \lower.1ex \hbox{\it e}\,}}
\def\Im{{\cal I \mskip-5mu \lower.1ex \hbox{\it m}\,}}

\def\tev{\,{\ifmmode\mathrm {TeV}\else TeV\fi}}
\def\gev{\,{\ifmmode\mathrm {GeV}\else GeV\fi}}
\def\mev{\,{\ifmmode\mathrm {MeV}\else MeV\fi}}

\begin{document}

\title{ Neutron decay into a Dark Sector via Leptoquarks  }

\author{Sara Khatibi\orcidB{}}
\email{sara.khatibi@ut.ac.ir}

\affiliation{ Department of Physics, University of Tehran, North Karegar Ave., Tehran 14395-547, Iran }


\begin{abstract}
\label{abstract}
In this paper, we extend the Standard Model (SM) scalar sector with scalar leptoquarks (LQ) 
as a portal to the dark sector to resolve some observational anomalies simultaneously. 
We introduce LQ coupling to scalar dark matter (DM) to suggest an exotic decay channel for the neutron into scalar DM and an SM anti-neutrino. 
If the branching ratio of this new neutron decay channel is $1\%$,  a long-standing discrepancy 
in the measured neutron lifetime between two different experimental methods, bottle and beam experiments, can be solved.
The mass of the scalar DM produced from neutron decay should be in a narrow range and as a result, 
its production in the early universe is challenging.
We discuss that the freeze-in mechanism can produce this scalar DM in the early universe with the correct relic abundance.
Then we show that the model can explain other SM anomalies like the muon $(g-2)$, and $R_{D^{(*)}}$ anomaly
simultaneously.
\end{abstract}   


\maketitle


\newpage
\section{Introduction} \label{intro}
The Standard Model (SM) of particle physics is one of the most successful theories 
and almost all of its predictions are consistent with experimental results. 
However, there are some observations that the SM failed to explain.
One of the intriguing challenges in particle physics is the neutron lifetime anomaly. 
It is well-known that the neutron dominantly decays to a proton, 
an electron, and an anti-electron-neutrino ($\beta$ decay) in the SM framework.
The neutron lifetime has been measured by two different methods in experiment, bottle and beam experiments.
In the bottle experiment, the ultra-cold neutrons are kept in a container for a time longer than the neutron lifetime, 
then the remaining neutrons are counted and the neutron lifetime is extracted.
The average neutron lifetime from five bottle experiments  
is~\cite{Mampe:1993an,Serebrov:2004zf,Pichlmaier:2010zz,Steyerl:2012zz,Arzumanov:2015tea},
\beq
\tau_n^{\text {bottle }}=879.6 \pm 0.6 \mathrm{~s}.
\eeq
In the second method, the beam experiment, the numbers of the produced protons from the neutron decay are counted 
and then the neutron lifetime is measured.
The average neutron lifetime from two beam experiments is longer than those 
from the previous method~\cite{Byrne:1996zz,Yue:2013qrc},
\beq
\tau_n^{\text {beam }}=888.0 \pm 2.0 \mathrm{~s}.
\eeq
There is a $4\sigma$ discrepancy in neutron lifetime measurements.
This discrepancy can be solved if the neutron partially decays to the invisible, for instance, particles in the dark sector 
(with a branching ratio around $1\%$)~\cite{Fornal:2018eol,Davoudiasl:2014gfa,Barducci:2018rlx,Ivanov:2018uuk,Cline:2018ami,Elahi:2020urr}.

On the other hand, numerous observations from galactic to cosmic scales indicate the existence of 
dark matter (DM) that corresponds to approximately $25\%$ of the energy budget of the Universe. 
Understanding the nature of DM is one of the longstanding problems in particle physics. 
Although lots of efforts have been down to unveil the DM nature, its properties are still unknown. 
For example, we don't know if the DM is a fermion or a boson,  how it interacts (non-gravitationally) 
with the SM particles, how it was produced in the early universe and the DM mass value 
(since the wide range of mass is still valid for the DM particle).

It is well-known that the leptoquark (LQ) models are the economical method to address most of the SM anomalies. 
The LQs can be a scalar or vector and can simultaneously couple to a quark and a lepton. 
In this paper, the scalar sector of the SM is extended by two scalar LQs ($S^{\alpha}_1$ and $S^{\beta}_1$) 
where both have the same quantum number under the SM gauge group but have different baryon and lepton numbers. 
Also, we add a dark scalar ($\phi$) which is a singlet under the SM. 
It is shown that by introducing a new coupling between the LQs and the dark scalar, 
which is a portal between the SM and the dark sectors, the neutron can decay to a dark scalar 
and an SM anti-neutrino through the scalar LQ mediators. 
Then it is indicated that the neutron lifetime anomaly can be evaded in a suitable parameter space region.

Since there are severe constraints on the baryon number violation process, 
we consider the new exotic neutron decay channel with respect to the baryon 
number~\cite{Super-Kamiokande:2016exg,Phillips:2014fgb,LHCb:2017vth,SNO:2017pha}. 
So, the new dark scalar carries the baryon and lepton numbers.
Furthermore, we show that the dark scalar can be a good DM candidate since it is the lightest particle with the baryon number in the model. 
On the other hand, the exotic neutron decay channel should be kinematically allowed 
and at the same time the proton decay should be prevented, so the mass of the scalar DM should be in the narrow range.
As a result, the production of such a scalar DM is challenging. However, we show that through the freeze-in scenario, 
it can be produced in the early universe and its relic density is compatible with the observed abundance of the DM. 

In the rest of the paper, we examine other SM anomalies that can be addressed by our model.
One of the established anomalies in the SM is related to the high-precision measurement of the magnetic moment of the muon. 
The SM prediction for the magnetic moment of the muon has around $5.1\sigma$ deviation from the combined 
result measurement from Brookhaven National Laboratory (BNL) and Fermi National Accelerator 
Laboratory (FNAL)~\cite{Aoyama:2020ynm,Muong-2:2006rrc,Muong-2:2021ojo,Muong-2:2021vma,Muong-2:2023cdq}.
To alleviate this problem we need new physics with extra particles.
Furthermore, the semi-leptonic decays of B-mesons are sensitive to new physics.
For example, the BaBar~\cite{BaBar:2012obs,BaBar:2013mob}, 
Belle~\cite{Belle:2015qfa,Belle:2016dyj,Belle:2017ilt,Belle:2019rba}, 
and LHCb~\cite{LHCb:2015gmp,LHCb:2017smo,LHCb:2017rln} experiments have measured the $R_D$ and $R_{D^{*}}$ observables 
where they have shown that their result has a deviation from the SM prediction. 
Although the current uncertainties should be understood better, one can study the new physics effects on these anomalies.
Then, we show that our model can solve these SM anomalies simultaneously.
It is worth noting that our paper differs from previous studies in the following ways:
\begin{itemize}
\item In previous papers, the DM candidate was a fermion particle, but in this paper, the neutron decays to scalar DM.

\item We utilized LQ particles as mediator particles due to their intriguing phenomenology and ability 
to explain multiple anomalies at once.

\item We employed the Freeze-in production mechanism during the early universe to account for the DM relic density.

\item In this proposed final state of neutron exotic decay, a SM neutrino is produced. 
Despite their weak interactions, the existence of the neutrinos in the final state could yield a 
distinct signature compared to previous proposals.
\end{itemize}
 
The organization of the paper is as follows. 
In Section~\ref{sec:model}, we explain the model in detail. In section~\ref{sec:Pheno} the different 
phenomenological aspects of the model are discussed. Finally, section~\ref{sec:summary} summarises the paper.
\section{The Model}\label{sec:model}
We extend the SM scalar sector by three new particles. The $S^{\alpha}_1$ and $S^{\beta}_1$ are the LQ scalars and
have the same quantum number under the SM gauge group $(\overline{\mathbf{3}}, \mathbf{1}, 1 / 3)$, 
however, they have different baryon and lepton numbers.
The third scalar $\phi$ is singlet under the SM gauge symmetries but it carries the baryon and lepton numbers.
Table~\ref{tab:Particle} represents all the new scalars with their quantum numbers.
\begin{table}[t!]
\begin{center}
\begin{tabular}{|c cc  cc cc|}
\hline 
Particles && B && L &&$SU(3)_{C}\times SU(2)_{_L} \times U(1)_{Y}  $ \\
\hline
\hline
$S^{\alpha}_1$&& $-1/3$ && $-1$  &&($\overline{\mathbf{3}}$~~~~~,~~~1~~~,~~~~~1/3)\\
$S^{\beta}_1$ && $2/3$ && $0$   && ($\overline{\mathbf{3}}$~~~~~,~~~1~~~,~~~~~1/3)  \\
$\phi$ && $1$ &&  $1$    &&(1~~~~~,~~~1~~~~,~~~~~0)  \\
\hline
\end{tabular}
\end{center}
\caption{The quantum numbers of the new scalars. The second and third columns show the baryon and lepton numbers, respectively. 
The last column presents the quantum numbers under the SM gauge groups. }
\label{tab:Particle}
\end{table}%
The Lagrangian of the model contains all the particle interactions, and can be written as follows,
\beq
\cL=\cL_{_{\text{SM}}}+\cL_{_{\text{LQ, int}}}+\cL_{_{\text{Scalars}}},
\label{Eq:mainLag}
\eeq
where $\cL_{_{\rm{SM}}}$ is the SM Lagrangian and $\cL_{_{\rm{Scalars}}}$ contains 
all kinetic, mass and interaction terms for scalars. $\cL_{_{\rm{LQ, int}}}$ indicates the scalar LQ interactions with
the SM fermion fields and has the following form~\cite{Dorsner:2016wpm},
\beq
\cL_{_{\text{LQ, int}}} = y_{ i j}^{L L} \bar{Q}_L^{C i, a} S^{\alpha}_{1} \epsilon^{a b} L_L^{j, b}
+z_{ i j}^{L L} \bar{Q}_L^{C i, a} {S^{\beta}_{1}}^* \epsilon^{a b} Q_L^{j, b}
+ y_{ i j}^{R R} \bar{u}_R^{C i} S^{\alpha}_1 e_R^j+\text { h.c. },
\eeq 
where $Q_{L}~(L_{L})$ indicate the left-handed quark (lepton) doublet
and $u_{R}~(e_{R})$ show the right-handed up-type quark (charged lepton).
The flavor $(SU(2))$ indices are shown by $i,j =1,2,3~(a,b=1,2)$, 
and $\epsilon^{a b} =( i \sigma^2)^{a b}$ that $\sigma^2$ is the second Pauli matrix. 
For the fermion $\psi$, we use the following notation, $\bar{\psi}=\psi^{\dagger} \gamma^{0}$
and $\psi^{C}=C \bar{\psi}^{T}$, where $C= i \gamma^2 \gamma^0$ is the charge conjugation operator.
The $y^{L L}$ and $y^{R R}$ are  completely arbitrary $3 \times 3$ matrices 
but $z^{L L}$ is a symmetric matrix in flavor space $(z_{i j}^{L L}=z_{j i}^{L L})$.
After the contraction in the $SU(2)$ space, we have the following interaction terms for the LQ scalars with the SM fermions,
\begin{eqnarray}
\nonumber
\cL_{_{\text{LQ, int}} } = &-&\left(y^{L L} U\right)_{i j} \bar{d}_L^{C i}  S^{\alpha}_{1} \nu_L^j
+\left(V^T y^{L L}\right)_{i j} \bar{u}_L^{C i}  S^{\alpha}_{1} e_L^j
+\left(V^T z^{L L}\right)_{i j} \bar{u}_L^{C i} {S^{\beta}_{1}}^* d_L^j  \\ 
&-&\left(z^{L L} V^{\dagger}\right)_{i j} \bar{d}_L^{C i} {S^{\beta}_{1}}^* u_L^j 
+y_{i j}^{R R} \bar{u}_R^{C i} S^{\alpha}_1 e_R^j+\text { h.c.}~,
\label{Eq:lagLQ}
\end{eqnarray}
where $U$ is a Pontecorvo-Maki-Nakagawa-Sakata (PMNS) unitary mixing matrix 
and $V$ is a Cabibbo-Kobayashi-Maskawa (CKM) mixing matrix. All fields in the above equation are in the mass 
eigenstate basis.
Moreover, The scalar Lagrangian for a SM singlet scalar $\phi$ and two other scalars $S^{\alpha}_1$ 
and $S^{\alpha}_2$ is given by,
\begin{eqnarray}
\nonumber
\cL_{_{\text{Scalars}}} &=& \left|D_\mu \phi \right|^2- m_{_{\phi}}^2\left| \phi \right|^2 +\left|D_\mu S^{\alpha}_{1} \right|^2
-m_{_{S^{\alpha}_{1}}}^2\left|S^{\alpha}_{1}\right|^2
+\left|D_\mu S^{\beta}_{1} \right|^2
-m_{_{S^{\beta}_{1}}}^2\left|S^{\beta}_{1}\right|^2 
- \lambda_1\left|S^{\alpha}_1 \right|^4-\lambda_2\left|S^{\beta}_1 \right|^4\\ 
&-&\lambda_3\left|\phi \right|^4
-\lambda_4 |H|^2\left|S^{\alpha}_1 \right|^2-\lambda_5 |H|^2\left|S^{\beta}_1 \right|^2
-\lambda_6 |H|^2\left| \phi \right|^2   
-(\mu~S^{\alpha}_1 ({S^{\beta}_{1}})^* \phi+\text { h.c.}) , \nonumber
\label{Eq:lagscalar}
\end{eqnarray} 
where $H$ is the SM Higgs doublet. The $\lambda_{i}$s are the dimensionless couplings 
whereas the $\mu$ has a dimension of mass.
The last term in the above Lagrangian plays a crucial role in the model because it is a portal between the SM and the dark sectors. 
As we will explain in the next section the $\phi$ is the lightest particle with the baryon number in our model, 
as a result, it is stable and can be a good DM candidate.

It is worth mentioning that, in the rest of the paper, for sake of simplicity and in order to resolve some anomalies simultaneously, 
we consider the following economical flavor ansatz,
\beq
y_{12}^{LL} \neq 0,~~~~~~ y_{33}^{LL}\neq 0,~~~~~~ y_{32}^{LL} \neq 0,~~~~~~ y_{23}^{LL}\neq 0,~~~~~~y_{32}^{RR} \neq 0,~~~~~~z_{11}^{LL} \neq 0,
\label{ansatz}
\eeq
and other couplings in the LQ Lagrangian (Eq.~\ref{Eq:lagLQ}) are considered to be zero.

\section{Phenomenology}
\label{sec:Pheno}

In this section, we study different phenomenological aspects of our model. 
In the first subsection~\ref{sec:Neutron Decay}, we show how the neutron 
can decay to a scalar DM and an anti-neutrino via the scalar LQs. 
We find an appropriate benchmark in our model in which the neutron decay anomaly resolve. 
The DM production in the early universe and calculating DM relic abundance are presented in the subsection~\ref{sec:ProDM}.
Then in the subsection~\ref{sec:Muon}, we explain how our setup can eliminate 
the muon anomalous magnetic moment.
In subsection~\ref{sec:Banomaly}, we indicate that the $R_{D^{(*)}}$ anomaly can be alleviated 
through our model with the proper parameter space.

\subsection{Neutron Decay Anomaly}\label{sec:Neutron Decay}
\begin{figure}[t]
   \centering
   \begin{subfigure}[b]{0.4\textwidth}
       \includegraphics[width=5cm]{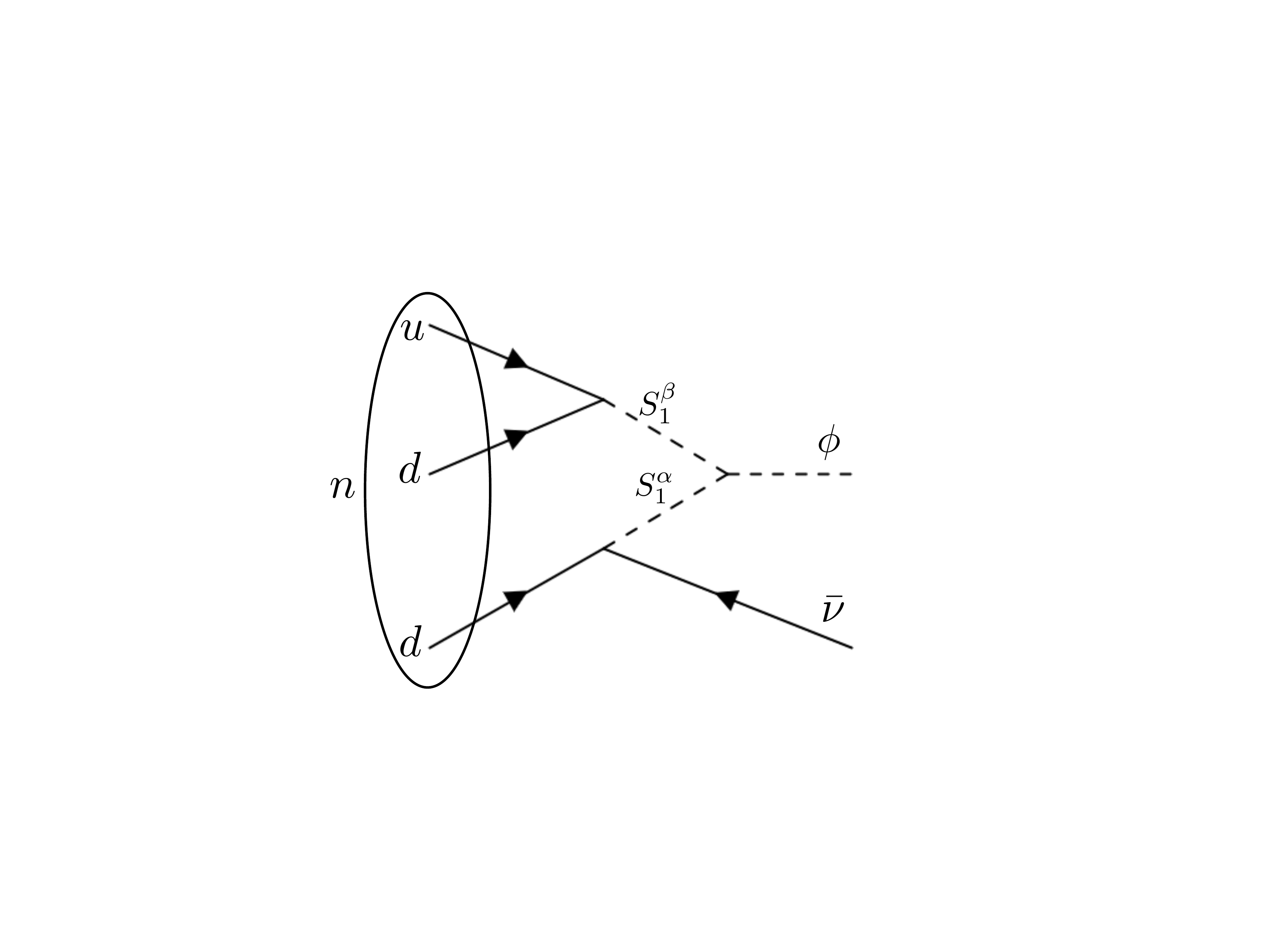}
       \caption{Neutron decay $n \rightarrow \phi  \bar{\nu}$}
       \label{Fig:dn}
   \end{subfigure}
   \hspace{2cm}
   \begin{subfigure}[b]{0.4\textwidth}
       \includegraphics[width=5cm]{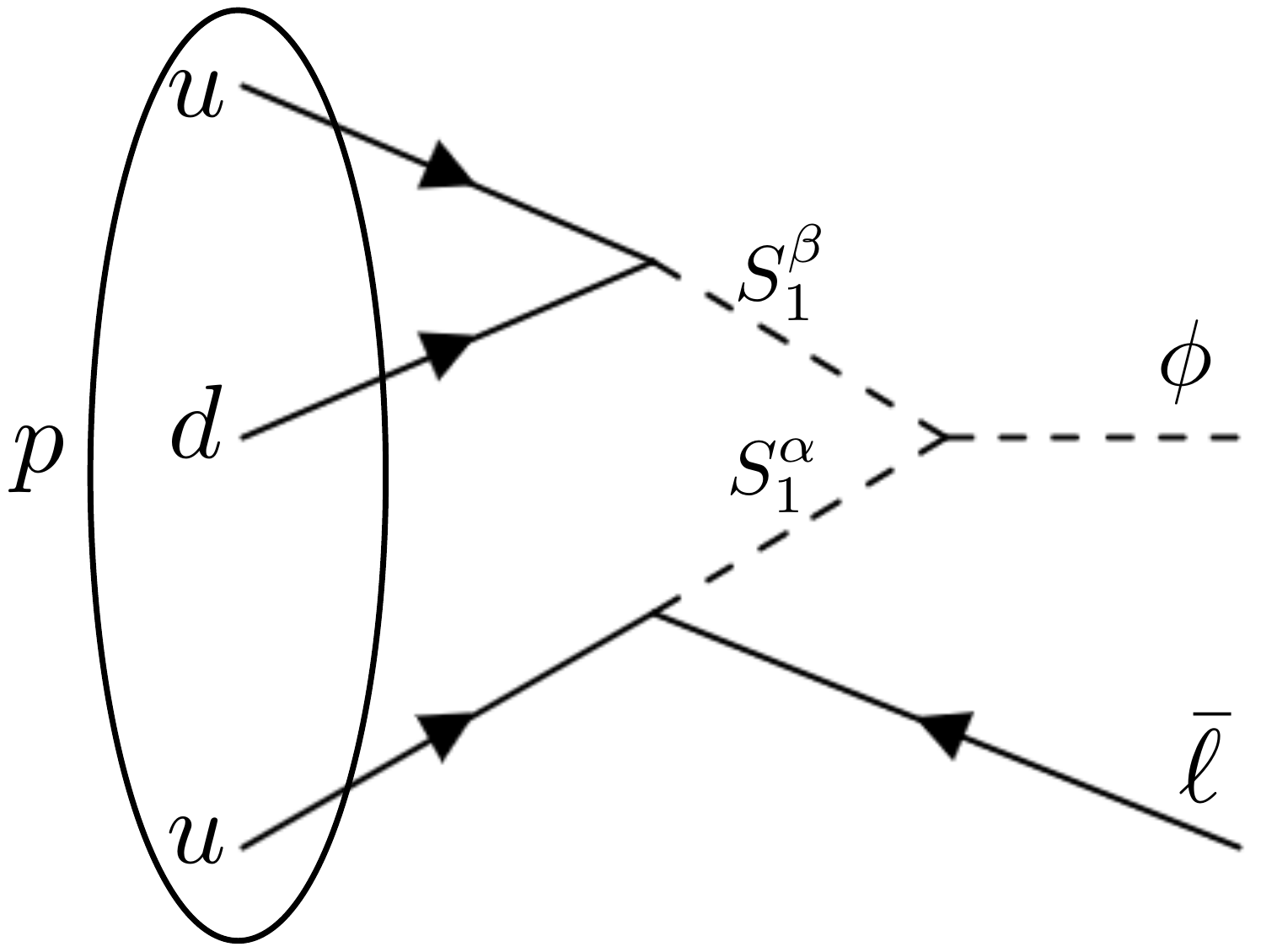}
       \caption{Proton decay $p \rightarrow \phi  \bar{\ell}$}
       \label{Fig:dp}
    \end{subfigure}
    \caption{The Feynman diagrams contributing to neutron and proton decay.}
    \label{Fig:decays}
\end{figure}

As we mentioned before, one of the intriguing challenges in particle physics is the neutron lifetime anomaly. 
In order to evade such an impasse the neutron can partially decay into the dark sector. 
In our model, the neutron can decay into a scalar DM and an SM anti-neutrino ($n \rightarrow \phi \bar{\nu}$). 
According to the Lagrangian in Eq.~\ref{Eq:mainLag} and considering the flavor ansatz in Eq.~\ref{ansatz}, 
the following terms have contribution to the exotic neutron decay,
\begin{eqnarray}
\cL_{n \rightarrow \phi \bar{\nu}^i} = - y_{12}^{L L} U_{2i}  \bar{d}_L^{C} \nu^i_{L} S^{\alpha}_1 
+ z_{11}^{L L} V_{1 1} \bar{u}_L^{C} d_L (S^{\beta}_1)^*
-\mu~S^{\alpha}_1 ({S^{\beta}_{1}})^* \phi + \text{h.c.},
\label{Eq:neutronLQ}
\end{eqnarray}
where $\nu^i_{L}$ can be any mass eigenstate of the SM neutrino.
The corresponding Feynman diagram for neutron decay to a scalar DM and an anti-neutrino is shown in Fig.~\ref{Fig:dn}.

It is worth mentioning that, there are some constraints on the scalar DM mass $(m_{_{\phi}})$. 
First, for the exotic neutron decay channel to be kinematically allowed, the $\phi$ should be lighter than the neutron.
The other bound comes from proton decay.
In our model, the proton also can decay to a scalar DM and an anti-muon (according to our flavor ansatz). 
The Feynman diagram for proton decay is illustrated in Fig.~\ref{Fig:dp},
and the following Lagrangian terms give rise to proton decays, 
\begin{eqnarray}
\cL_{p \rightarrow \phi \bar{\mu}} = + y_{12}^{L L} V_{11}  \bar{u}_L^{C} \mu_{L} S^{\alpha}_1 
+ z_{11}^{L L} V_{1 1} \bar{u}_L^{C} d_L (S^{\beta}_1)^* 
-\mu~S^{\alpha}_1 ({S^{\beta}_{1}})^* \phi+\text{h.c.} .
\end{eqnarray}
To prevent proton decay, the scalar DM mass should be in the following range,
\begin{align}
\nonumber
m_p - m_{\mu} <  m_{\phi} < m_n 
~~~~~~~\rightarrow~~~~~~~
832.71~\text{MeV}  <  m_{\phi} < 939.565~\text{MeV}.
\end{align}
Moreover, nuclear physics puts other bounds on the $m_{_{\phi}}$. 
The most stringent constraint is required to prevent nuclear decay of ${^9}\rm{Be}$~\cite{Fornal:2018eol},
\beq
937.900~\text{MeV} < m_{\phi} < 939.565 ~\text{MeV}.
\nonumber
\eeq
According to the aforementioned limits, we chose $m_{\phi}=938$ MeV as our benchmark. 
As a result, the $\phi$ is stable since it is the lightest particle with baryon number.

There are some constraints on the (first and second generation) scalar LQs mass from the CMS experiment, 
where they searched for the single and pair production of scalar LQ.
The current bounds require that the mass of the scalar LQ should be larger than 1.36 TeV~\cite{CMS:2018ncu,CMS:2018lab,CMS:2015xzc}.
So, the scalar LQs are heavier than other particles in the model and they can be integrated out from the Lagrangian. 
As a result, the effective Lagrangian contributing to the exotic neutron decay is given by, 
\beq
\cL_{n \rightarrow \phi \bar{\nu}^i}^{\mathrm{eff}}=\kappa^i \bar{n}^{C}_{L}  \nu^i_{L} \phi^*+ \text{h.c.},
\eeq
where
\beq
\kappa_i= \frac{\mu~\beta ( y_{12}^{L L} U_{2i})   (z_{11}^{L L} V_{1 1}) }{ m_{_{S^{\alpha}_{1}}}^2 m_{_{S^{\beta}_{1}}}^2},
\eeq
that $\beta \cong 0.014~\mathrm{GeV}^3$ form lattice QCD~\cite{Aoki:2017puj}.
According to the above effective Lagrangian, the exotic neutron decay width to a scalar DM and an anti-neutrino can be calculated as follows,
\beq
\Delta \Gamma(n\rightarrow\phi \bar{\nu})= \sum_{i} |{\kappa_i}|^2 \frac{1}{16 \pi m_n^3}(m_n^2- m_{_{\phi}}^2)^2,
\eeq
where
\beq
\sum_{i} |{\kappa_i}|^2= |\frac{\mu~\beta ( y_{12}^{L L})   (z_{11}^{L L} V_{1 1}) }{ m_{_{S^{\alpha}_{1}}}^2 m_{_{S^{\beta}_{1}}}^2}|^2,
\eeq
and the unitary condition of the PMNS matrix $(\sum_{i} |U_{2i}|^2=1)$ is used. 
To resolve the neutron decay anomaly, the exotic decay width should have the following value~\cite{Fornal:2018eol},
\beq
\Delta \Gamma(n\rightarrow\phi \bar{\nu})=\Gamma_n^{\text {bottle }}-\Gamma_n^{\text {beam }} 
\simeq 7.1 \times 10^{-30}~\mathrm{GeV}.
\eeq
Therefore, the following limit is imposed on the combination of the model parameters,
\beq
|\frac{\mu ~ y_{12}^{L L} z_{11}^{L L}  }{ m_{_{S^{\alpha}_{1}}}^2 m_{_{S^{\beta}_{1}}}^2}|^2 
\simeq 1.8 \times 10^{-19} ~\rm{GeV}^{-6}.
\label{Eq:limit}
\eeq
\begin{figure}[t!]
\centering
\includegraphics[height=6cm, width=6cm]{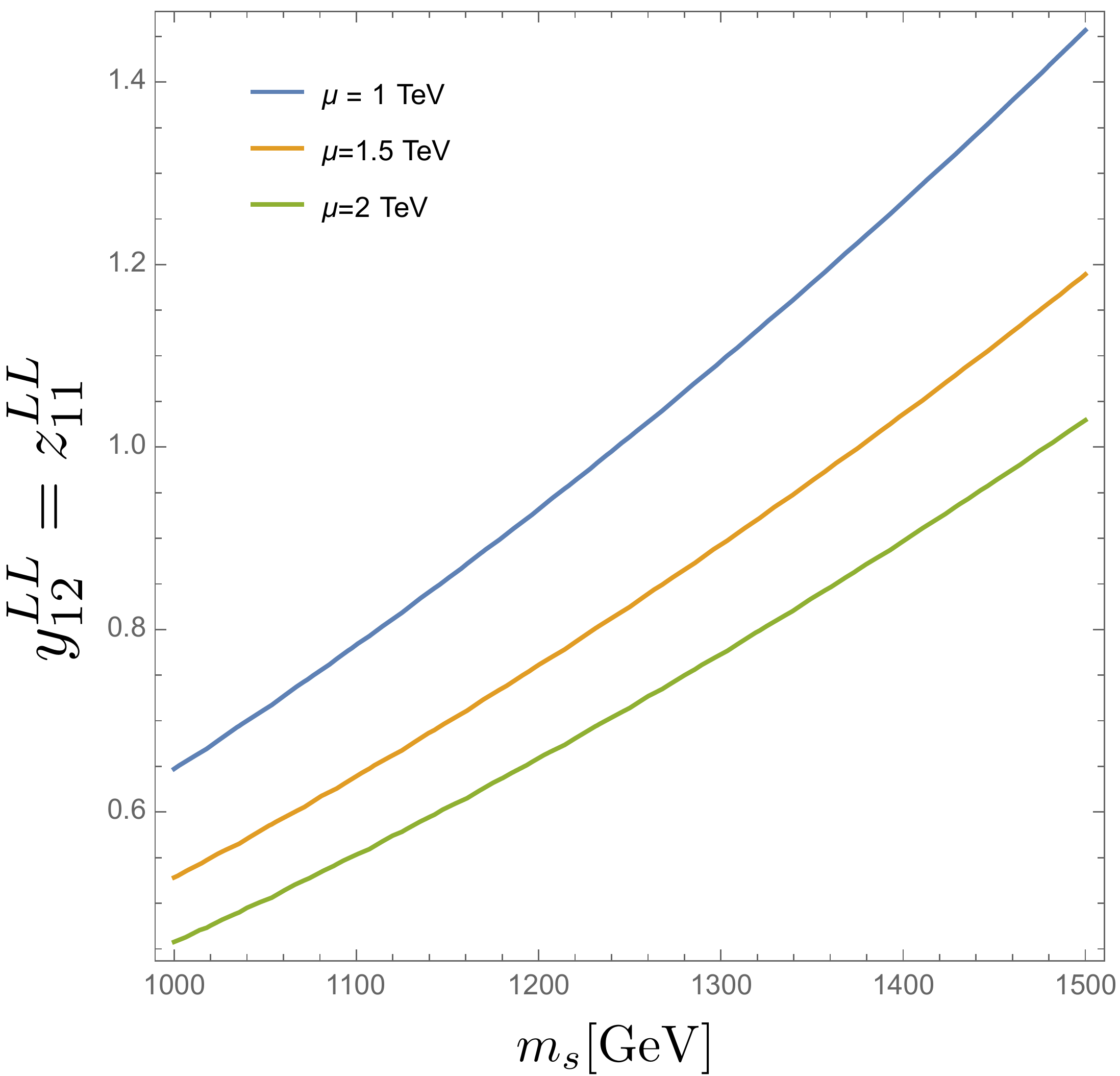}
\hspace{1.5cm}
\includegraphics[height=6cm, width=6cm]{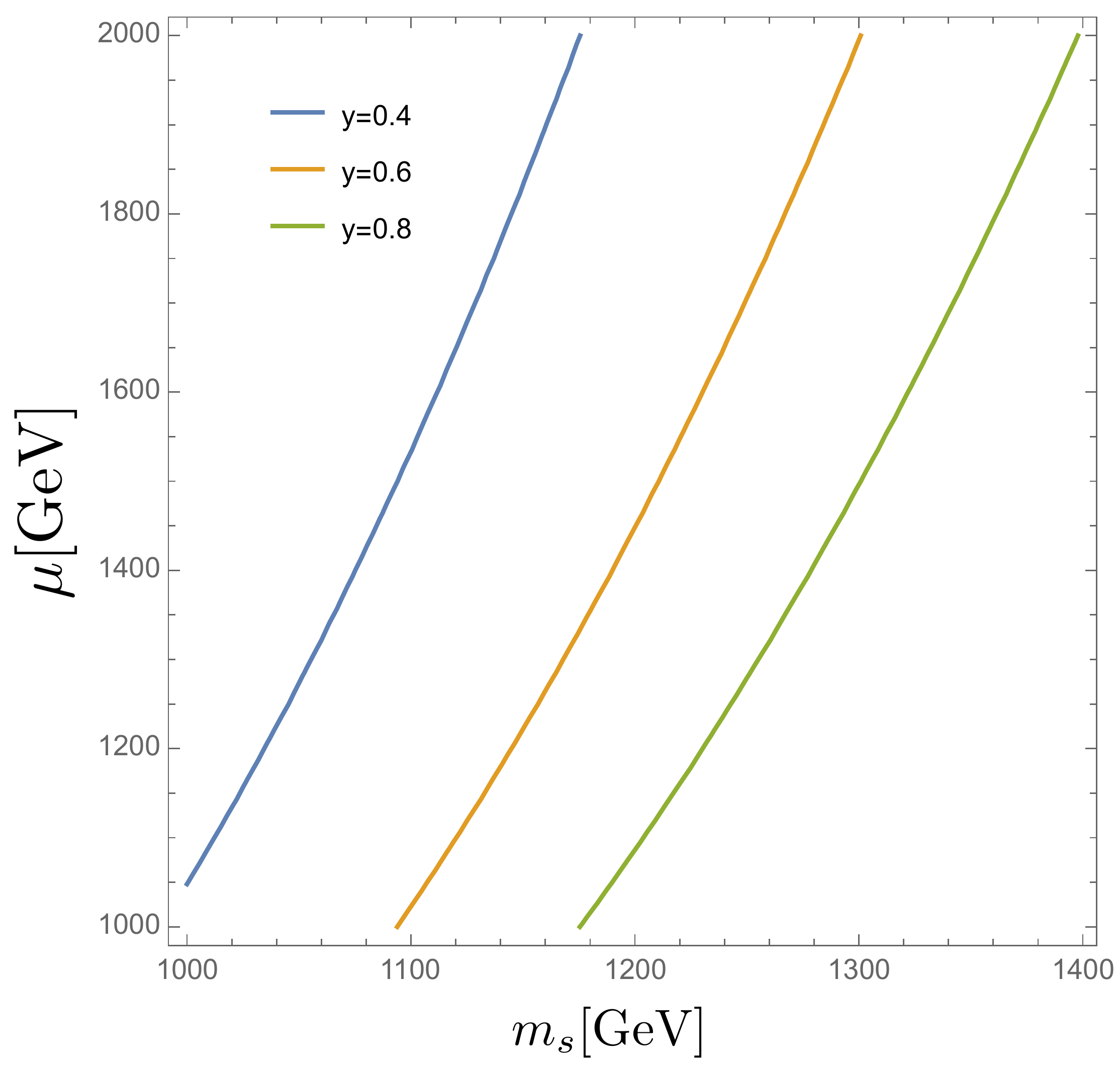}
\caption{The left panel shows the LQ coupling as a function of the LQ mass for $\mu=$ 1,1.5, and 2 TeV. 
The right panel shows the $\mu$ coupling as a function of the LQ mass for $y_{12}^{L L}= z_{11}^{L L}=$ 0.4,0.6, and 0.8. }
\label{Fig:neutronlimits}
\end{figure}
In Fig.~\ref{Fig:neutronlimits}, we show the accepted values for the LQ mass and LQ coupling with the SM fermion according to the above limit.
For simplicity, we assume that $m_{_{S^{\alpha}_{1}}}=m_{_{S^{\beta}_{1}}}$ and $y_{12}^{L L}= z_{11}^{L L}$.
The Left panel shows the LQ coupling as a function of the LQ mass for different values of $\mu$, and the right panel 
shows the dimensionful coupling $\mu$ as a function of the LQ mass for different values of LQ coupling.
According to the figures, the LQ mass around 1.36 TeV and $y_{12}^{L L}= z_{11}^{L L}=$ 0.8 are allowed with 
dimensionful coupling $\mu$ around the TeV scale. We chose these values as our benchmark.\footnote{It should be noted that the new scalar LQ 
($S^{\alpha}_1$ in Eq.~\ref{Eq:lagLQ}) can impact the decay of charged pion into muon and neutrino,
thereby affecting the following well-measured observable~\cite{10.1093/ptep/ptac097}:
\begin{equation}
\frac{\Gamma(\pi^- \rightarrow e^- \bar{\nu_e})}{\Gamma(\pi^- \rightarrow \mu^- \bar{\nu_{\mu}})} = 1.230(4) \times 10 ^{-4}. \nonumber
\end{equation}
We calculated the contribution of the new diagram channel and its interference with the SM channel. Based on our benchmark values, we found the resulting ratio to be:
\begin{equation}
\frac{\Gamma(\pi^- \rightarrow e^- \bar{\nu_e})}{\Gamma(\pi^- \rightarrow \mu^- \bar{\nu_{\mu}})} = \frac{\Gamma_{\rm{SM}}(\pi^- \rightarrow e^- \bar{\nu_e})}{\Gamma_{\rm{SM}}(\pi^- \rightarrow \mu^- \bar{\nu_{\mu}})+\Gamma_{\rm{int}}(\pi^-\rightarrow\mu^- \bar{\nu_{\mu}})+\Gamma_{\rm{exotic}}(\pi^-\rightarrow\mu^- \bar{\nu_{\mu}})}=1.2318\times10^{-4}, \nonumber
\end{equation}
which is consistent with the current measurements.}

It is worth mentioning that the neutron star (NS) can put constraints on the models that suggest a new neutron decay channel. 
The Tolman-Oppenheimer-Volko (TOV) equations determine the NS structure~\cite{Tolman:1939jz,Oppenheimer:1939ne}. 
If we integrate the TOV equations from the center of the NS, where the pressure is a constant, to the radius
of the NS with zero pressure, we can find the mass of NS as a function of its radius. Also, we can predict the maximum possible 
mass of the NS. However, to do the above procedure we need to have the Equation of State (EOS) of the NS. The EOS gives the 
relation between the energy density and the pressure for the NS.
The new neutron dark decay channel causes the DM to be thermalized inside the NS and as a result, the EOS would be changed.
Ref.~\cite{McKeen:2018xwc} showed that for a non-interacting DM with mass below the neutron mass (to have a kinematically 
allowed neutron decay channel), the DM produces more energy than the pressure and the EOS of the NS becomes softer. 
As a result, these models predict maximum mass for neutron stars below  $0.7 M_\odot\ $, which is in contradiction 
with observation. According to the data the current maximum mass for the NS is about $2 M_\odot\ $~\cite{Demorest:2010bx,Antoniadis:2013pzd}.

However, Ref.~\cite{McKeen:2018xwc,Ellis:2018bkr} showed that the DM model with mass greater than $1.2$ GeV or 
the repulsive self-interacting DM model can escape from these constraints.
For instance, if the DM is charged under a new gauge symmetry (U(1) or SU(2)), NS limits can be 
evaded for a suitable value of dark gauge mediator mass and gauge coupling~\cite{Cline:2018ami,Elahi:2020urr}. 
In our model, the neutron decays to a dark scalar and an anti-neutrino $(n \rightarrow \phi \bar{\nu})$. 
According to the Lagrangian in Eq.~\ref{Eq:lagscalar}, the scalar DM $(\phi)$ has a repulsive self-interaction 
term $\lambda_3\left|\phi \right|^4$ for the $\lambda_3>0$. As a result, we can evade NS constraints.

\subsection{DM Production}
\label{sec:ProDM}

In our model, the DM can be produced through the freeze-in 
mechanism in early universe~\cite{Hall:2009bx,Elahi:2014fsa}. 
In this mechanism, the DM has negligible abundance at the early time, however, some interaction with bath particles can produce the DM. 
In our case, after the QCD phase transition, the neutron and anti-neutron can decay into $\phi$ 
and contribute to the DM relic density. 
Although this contribution is negligible since the obtained relic density for $\phi$ from the neutron decay 
is four orders of magnitude less than the observed cosmological DM relic~\cite{Strumia:2021ybk}. 
Another type of interaction that can contribute to $\phi$ relic abundance 
is $n \pi^0 \rightarrow \phi \bar{\nu}$ scattering.
The number density of the DM ($n_{\phi}$) can be calculated by the Boltzmann equation in the freeze-in scenario,
\begin{eqnarray}
\dot{n}_{\phi} +3 n_{\phi} H \approx \int d \Pi_{n} d \Pi_{\pi} 
d \Pi_{\bar{\nu}} d \Pi_{\phi}(2 \pi)^4 
 \delta^4\left(p_{n}+p_{\pi}-p_{\bar{\nu}}-p_{\phi}\right)
 |M|_{n \pi \rightarrow \bar{\nu} {\phi}}^2 f_{n} f_{\pi},
\end{eqnarray}
where the $H$ is the Hubble parameter, 
$ d \Pi_i=d^3 p_i /(2 \pi)^3 2 E_i$ are phase space elements and 
$f_{i}=\left(e^{E_{i} / T} \pm 1\right)^{-1} $ are phase space densities.
Assuming the initial particles are in thermal equilibrium we can consider 
$f_{i} \approx e^{-E_{i} / T}$,  
and the Boltzmann equation can have the following form~\cite{Edsjo:1997bg},
\begin{eqnarray}
\dot{n}_{\phi} +3 n_{\phi} H \approx \frac{ T}{512 \pi^6} \int_{(m_n+m_{\pi})^2}^{\infty} ds~ d\Omega ~ 
P_{B_1 B_2}~P_{B_3 {\phi}}~|M|_{n \pi \rightarrow \bar{\nu} {\phi}}^2 K_1(\sqrt{s} / T) / \sqrt{s},
\label{Eq:Bolt}
\end{eqnarray}
where the $s$ and $T$ are the center of mass energy of the interaction 
and temperature, respectively. 
The $K_1$ is the first modified Bessel Function of the 2nd kind, and
\begin{eqnarray}
P_{i j} \equiv \frac{\left[s-\left(m_i+m_j\right)^2\right]^{1 / 2}\left[s-\left(m_i-m_j\right)^2\right]^{1 / 2}}{2 \sqrt{s}}.
\end{eqnarray}
The angular integration over the squared amplitude for 
$n \pi \rightarrow \bar{\nu} \phi$ interaction is as follows,
\beq
\int d\Omega~|M|_{n \pi \rightarrow \bar{\nu} {\phi}}^2 = 4 \pi \lambda^2 
\frac{(s- m^2_{\phi})(s+m^2_n-m^2_{\pi})}{2s}, 
\eeq
where $\lambda^2=|\frac{\mu~\beta ~ y_{12}^{L L} z_{11}^{L L} g^2_s }{ m_{_{S^{\alpha}_{1}}}^2 m_{_{S^{\beta}_{1}}}^2}|^2$ 
is the effective coupling and $g_s$ is the strong coupling constant.
If we use the yield definition, $Y_{\phi} \equiv n_{\phi}/S$ where $S$ is the entropy density, and consider
$\dot{T}=-HT$ the left part of Eq.~\ref{Eq:Bolt} becomes,
\beq
\dot{n}_{\phi}+3 n_{\phi} H = -S H T \frac{dY_{\phi}}{dT},
\eeq
where $S=2 \pi^2 g_*^S T^3 / 45$, $H=1.66 \sqrt{g_*^\rho} T^2 / M_{P l}$, and 
$M_{P l}$ is the non-reduced Planck mass.
The $g_*^{S,\rho}$ are the effective numbers of degrees of freedom in the bath at the freeze-in temperature 
for the entropy and energy density, respectively.
And finally, the variation of yield is given by,
\beq
\frac{dY_{\phi}}{dT} \approx  \frac{-1}{SHT} 
\frac{   4 \pi \lambda^2 T}{512 \pi^6}  \int_{(m_n+m_{\pi})^2}^{\infty} 
\frac{\sqrt{s-(m_n+m_{\pi})^2}}{2\sqrt{s}}\frac{s-m^2_{\phi}}{2\sqrt{s}} 
\frac{(s- m^2_{\phi})(s+m^2_n-m^2_{\pi})}{2s}
\frac{ K_1(\sqrt{s} / T)}{\sqrt{s}}~d s  .
\eeq
By doing the temperature integral with $T_{\rm{min}}= T_{\rm{BBN}}=1~\rm{MeV}$ and 
$T_{\rm{max}} = \Lambda_{\rm{QCD}}=180~\rm{MeV}$, we can obtain the yield of the DM at present $(Y^0_{\phi})$. 
In this temperature range, the $g_*^{S,\rho}$ is $17.25$. 
Then the DM relic density can be calculated by the following formula, 
\beq
\Omega_{\phi} h^2=\frac{m_{\phi} Y^0_{\phi} S_0}{\rho_c / h^2}, 
\eeq
where $S_0=2890 / \mathrm{cm}^3$ is the entropy density at the present time 
and $\rho_c / h^2=1.05 \times 10^{-5} \rm{GeV}/\rm{cm}^3$ that $\rho_c$ is the critical density. 
As we mentioned before, the model parameters are constrained by the neutron lifetime anomaly (Eq.~\ref{Eq:limit}),
so the value for $\lambda^2=|\frac{\mu~\beta ~ y_{12}^{L L} z_{11}^{L L} g^2_s }{ m_{_{S^{\alpha}_{1}}}^2 m_{_{S^{\beta}_{1}}}^2}|^2$ 
is fixed. Therefore, the DM relic density in the model is given by, 
\beq
\Omega_{\phi} h^2 \approx 0.12 (\frac{\lambda}{7.38 \times 10^{-11}})^2, 
\eeq
which is consistent with the Planck collaboration report ($\Omega_{\rm{DM}} h^2=0.12$)~\cite{Planck:2018vyg}. 
It is also noteworthy to mention that, we do a naive calculation here. For more accuracy, someone should consider the following points:
\begin{itemize}
\item Some other similar processes can contribute to the $\phi$ relic density such as $p \pi^0 \rightarrow \phi \bar{\mu}$. 
However, their contributions to the DM abundance should be in the same order as the $n \pi^0 \rightarrow \phi \bar{\nu}$ process. 
\item  As reviewed above, the DM yield is strongly dependent on the QCD confinement scale $(T_{\rm{max}} = \Lambda_{\rm{QCD}})$ 
and the value of the strong coupling constant ($g_s$).
Since there is a wide range for $\Lambda_{\rm{QCD}}$ from 100 MeV to 1 GeV, the value of the $Y^0_{\phi}$ is changing dramatically in this range.
\item The effect of the pion structure should also be considered in the effective coupling mentioned above ($\lambda$). 
Therefore, an extra factor (for example, the pion form factor) should multiply in the $\lambda$, 
which can reduce the value of the effective coupling by one or two orders of magnitude and change the DM relic density.
\end{itemize} 
However, by considering all of the above-mentioned points, in the worst-case scenario, the $\phi$ scalar can at least 
contribute to the $10 \%$ of the total DM abundance.

\subsection{Muon $g-2$}\label{sec:Muon}                                       

Another long-standing challenge in particle physics is the muon's anomalous magnetic moment.
The updated new world average from Brookhaven National Laboratory~\cite{Muong-2:2006rrc} 
and Fermi National Accelerator Laboratory~\cite{Muong-2:2021ojo,Muong-2:2021vma,Muong-2:2023cdq}
for $a_\mu= (g-2)_\mu /2$ has $5.1\sigma$ deviation from its SM prediction~\cite{Aoyama:2020ynm},
\beq
\delta a_\mu=a_\mu^{\mathrm{Exp} }-a_\mu^{\mathrm{SM}}=(249 \pm 48) \times 10^{-11}.
\label{Eq:gmuon}
\eeq
\begin{figure}[t!]
\centering
\includegraphics[height=3cm, width=5cm]{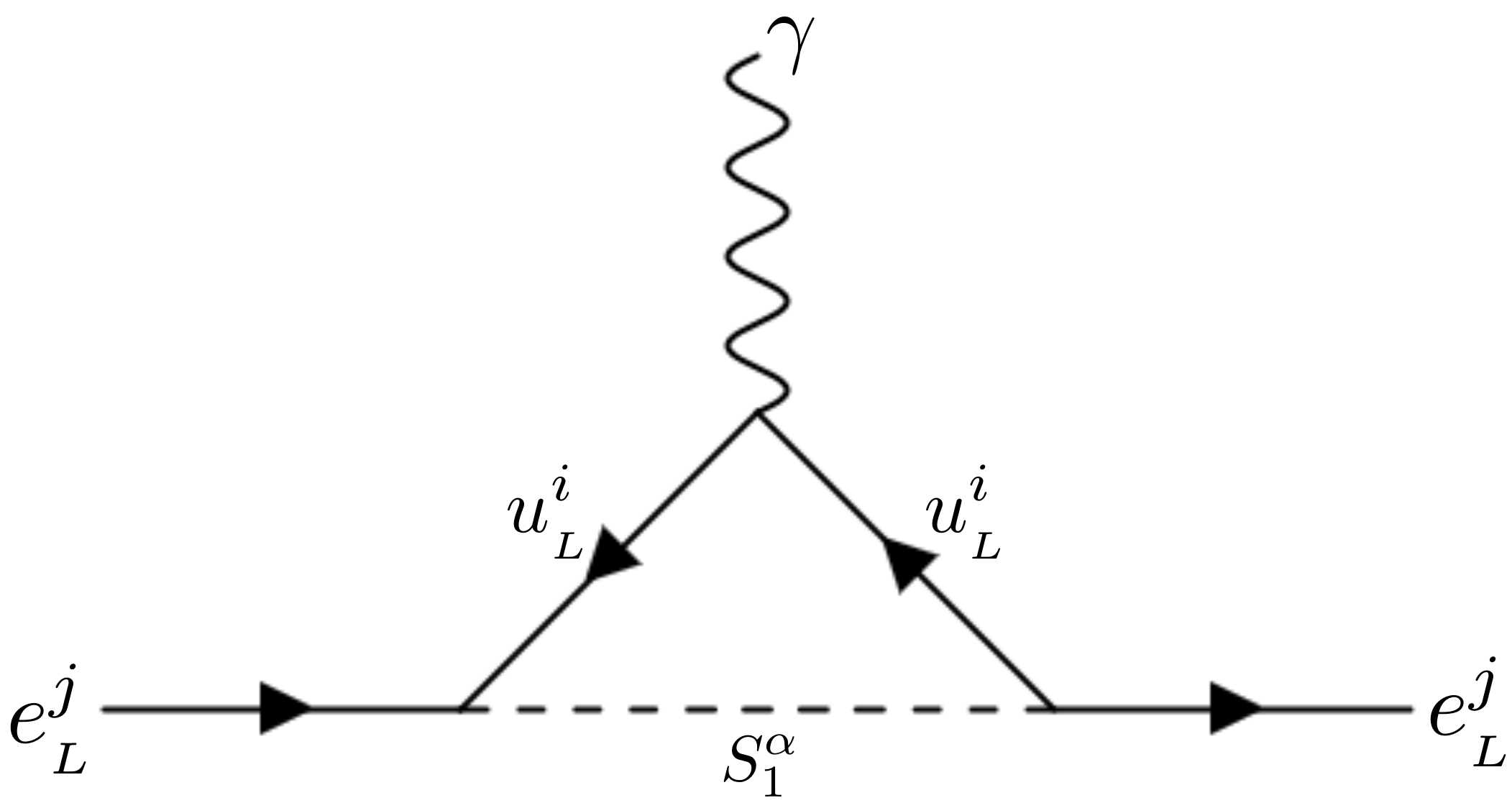}
\caption{The Feynman diagram contributes to the anomalous magnetic moment of the muon. }
\label{Fig:feynmuong-2}
\end{figure}
It is well-known that the scalar LQ can explain this 
anomaly~\cite{Dorsner:2016wpm,Djouadi:1989md,Dorsner:2019itg,Crivellin:2017zlb,Choi:2018stw,Cheung:2001ip,
ColuccioLeskow:2016dox,Crivellin:2020tsz,Queiroz:2014pra}.
In our setup, the scalar $S_1^{\alpha}$ can contribute to the magnetic moment of the muon and the relevant Feynman diagram is shown in Fig.~\ref{Fig:feynmuong-2}.
The related terms from the Lagrangian Eq.~\ref{Eq:lagLQ} are as follows,
\beq
\cL \supset \left(V^T y^{L L}\right)_{i j} \bar{u}_L^{C i}  S^{\alpha}_{1} e_L^j
+y_{i j}^{R R} \bar{u}_R^{C i} S^{\alpha}_1 e_R^j+ \text{h.c.},
\eeq
where the $u^i$ is the up-type quark ($u,c,t$) and $e^j$ is the charged lepton.
According to our economical ansatz (Eq.~\ref{ansatz}) and because of the large mass of the top quark
the following terms involving the top quark and muon have important effects on the $a_\mu$,
\beq
\cL \supset y_{32}^{L L} V_{33}  \bar{t}_L^{C} \mu_{L} S^{\alpha}_1 
+ y_{32}^{RR} \bar{t}_R^{C} \mu_R S^{\alpha}_1+ \text{h.c.} .
\eeq
The contribution of the above terms to the anomalous magnetic moment of the muon 
is given by~\cite{Djouadi:1989md},
\begin{eqnarray}
\delta a_\mu =-\frac{N_c m_\mu}{8 \pi^2  m_{_{S^{\alpha}_{1}}}^2 } [m_\mu(| y_{32}^{L L} V_{33}|^2
+| y_{32}^{RR}|^2) \mathcal{F}(x_t)
+m_t \operatorname{Re} [ (y_{32}^{RR})^ * (y_{32}^{L L} V_{33}) ] \mathcal{G}(x_t)],
 \label{muonmoment}
\end{eqnarray}
where $m_{\mu}$ and $m_{t}$ indicate the muon and top quark masses, respectively, $x_{t}=m_t^2/m_{_{S^{\alpha}_{1}}}^2$, 
and $N_c=3$ is the number of the QCD colors. The definition of $\mathcal{F}(x)$ and $\mathcal{G}(x)$ functions are, 
\begin{eqnarray}
\nonumber
& \mathcal{F}(x)=\frac{1}{3} f_S(x)-f_F(x,) \\
& \mathcal{G}(x)=\frac{1}{3} g_S(x)-g_F(x), 
\end{eqnarray}
where
\begin{eqnarray}
\nonumber
&f_S(x)=\frac{x+1}{4(1-x)^2}+\frac{x \log x}{2(1-x)^3}, ~~~~~ &g_S(x)=\frac{1}{x-1}-\frac{\log x}{(x-1)^2}, \\  \nonumber
&f_F(x)=\frac{x^2-5 x-2}{12(x-1)^3}+\frac{x \log x}{2(x-1)^4},  ~~~~~ &g_F(x)=\frac{x-3}{2(x-1)^2}+\frac{\log x}{(x-1)^3}.
\end{eqnarray}
\begin{figure}[t!]
\centering
\includegraphics[height=4.5cm, width=6.5cm]{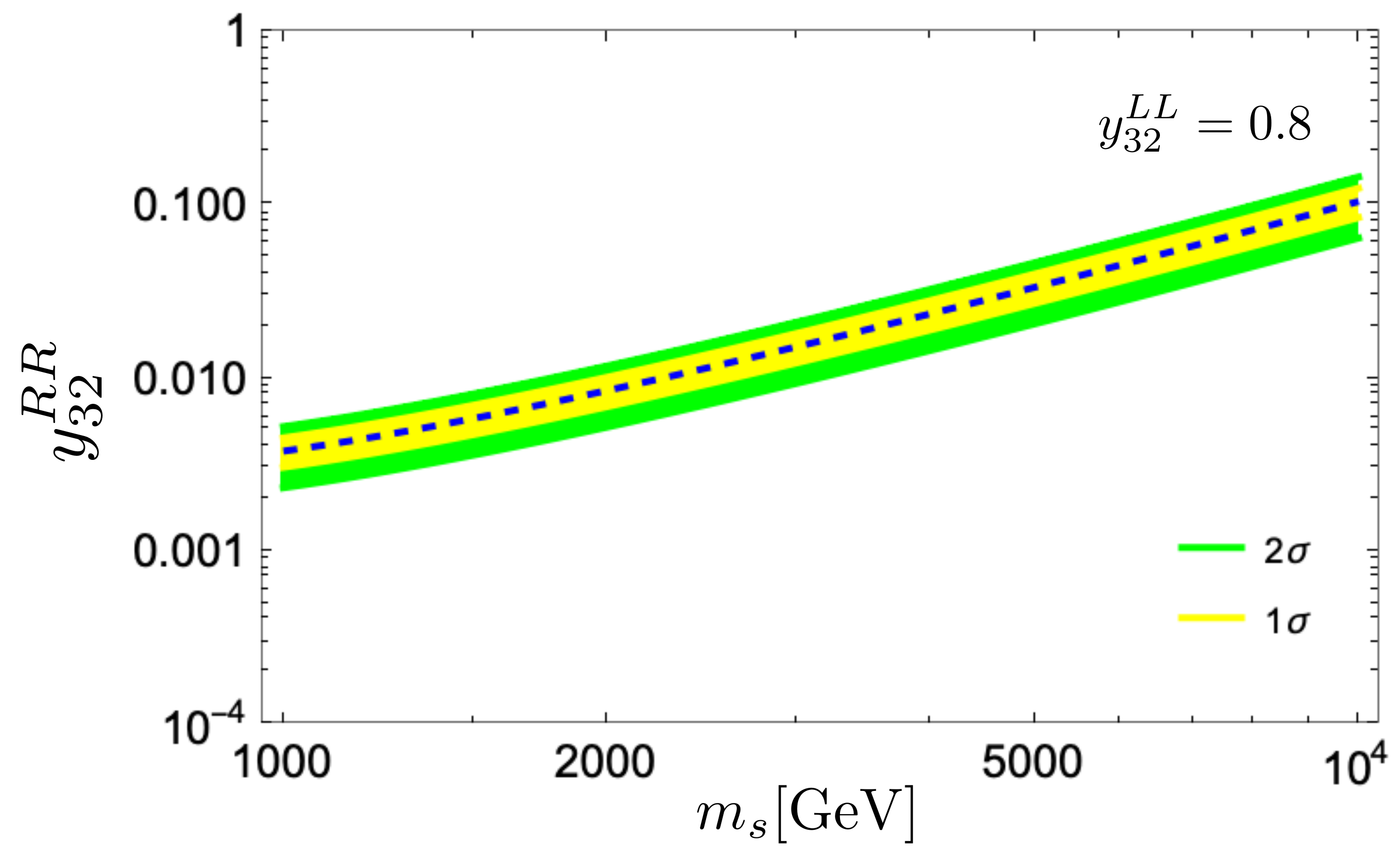}
\caption{The parameter space of the model ($y_{32}^{RR}$ coupling as a function of LQ's mass) explains the anomalous magnetic moment of the muon. The yellow and green regions indicate the 1$\sigma$ and 2$\sigma$ levels, respectively. The $y_{32}^{LL}$ value is fixed at 0.8.}
\label{Fig:limitmuon}
\end{figure}
As we can see, the first term in Eq.~\ref{muonmoment} is suppressed by muon mass.
The scalar LQ $(S_1^{\alpha})$ should have both left-handed
and right-handed couplings to generate the second term, which is proportional to top quark mass.
As a result of this chirality-enhanced effect and top quark mass, 
the significant contribution to the $a_{\mu}$ is as follows~\cite{Crivellin:2017zlb},
\begin{eqnarray}
\delta a_{\mu}  \approx  -\frac{N_c}{48 \pi^2 m_{_{S^{\alpha}_{1}}}^2  } m_{\mu}m_t 
\operatorname{Re}\left [ (y_{32}^{RR})^ * (y_{32}^{L L} V_{33}) \right]  
\left(7+4 \log \left(\frac{m_t^2}{m_{_{S^{\alpha}_{1}}}^2  } \right)\right).
\end{eqnarray}

Fig.~\ref{Fig:limitmuon} displays the parameter space ($y_{32}^{RR}$ coupling as a function of LQ's mass) 
where the model can account for the muon $g-2$ anomaly.
In this plot, $y_{32}^{LL}$ is fixed at 0.8, based on our benchmark from the neutron lifetime anomaly.

\subsection{$R_{D^{(*)}}$ Anomaly} \label{sec:Banomaly}
The semi-leptonic decays of B-mesons are sensitive to new physics.
The BaBar~\cite{BaBar:2012obs,BaBar:2013mob}, 
Belle~\cite{Belle:2015qfa,Belle:2016dyj,Belle:2017ilt,Belle:2019rba}, 
and LHCb~\cite{LHCb:2015gmp,LHCb:2017smo,LHCb:2017rln} experiments have measured the $R_D$ and $R_{D^{*}}$ observables 
where they have shown that their result has a deviation from the SM prediction. 
Although the current uncertainties should be understood better, one can study the new physics effects on these anomalies.
The definition of two anomalous observables are as follows,
\begin{eqnarray}
\nonumber
R_{D} &=& \frac{\rm{BR}(B\rightarrow D \tau \bar{\nu})}{\rm{BR}(B\rightarrow D \ell \bar{\nu})}, \\
R_{D^*} &=& \frac{\rm{BR}(B\rightarrow D^* \tau \bar{\nu})}{\rm{BR}(B\rightarrow D^* \ell \bar{\nu})},
\end{eqnarray}
where $\ell= e, \mu$ for BaBar and Bell and $\ell=\mu$ for LHCb.
The experimental world averages reporting by Heavy Flavor Averaging Group are~\cite{HFLAV:2022pwe},
\begin{eqnarray}
\nonumber
R^{\rm{exp}}_{D} &=& 0.356 \pm 0.029, \\
R^{\rm{exp}}_{D^*} &=& 0.284 \pm 0.013.
\end{eqnarray}
While the SM predictions for these observables are~\cite{HFLAV:2019otj},
\begin{eqnarray}
\nonumber
R^{\rm{SM}}_{D} &=& 0.298 \pm 0.004, \\
R^{\rm{SM}}_{D^*} &=& 0.254 \pm 0.005. 
\end{eqnarray}
The combination of the experimental result for $R_{D}$ and $R_{D^*}$ 
has a deviation from the SM prediction by about $3 \sigma$~\cite{WinNT}.
The effective Lagrangian for $b \rightarrow c \tau \nu^i$ is as follows,
\beq
\cL^{\rm{eff}}_{b \rightarrow c \tau \nu^i} = 
 -\frac{4 G_F}{\sqrt{2}} V_{cb} C_{cb} 
[(\bar{c}_{L} \gamma^{\mu} b_{L})(\bar{\tau}_{L}\gamma_{\mu} \nu^i_{L})]+ \text{h.c.},
\eeq
where $G_F$ is the Fermi constant and $C_{cb}=1$ in the SM. 
The new physics can contribute in the above effective operator. 

\begin{figure}[t!]
   \centering
       \includegraphics[width=5cm]{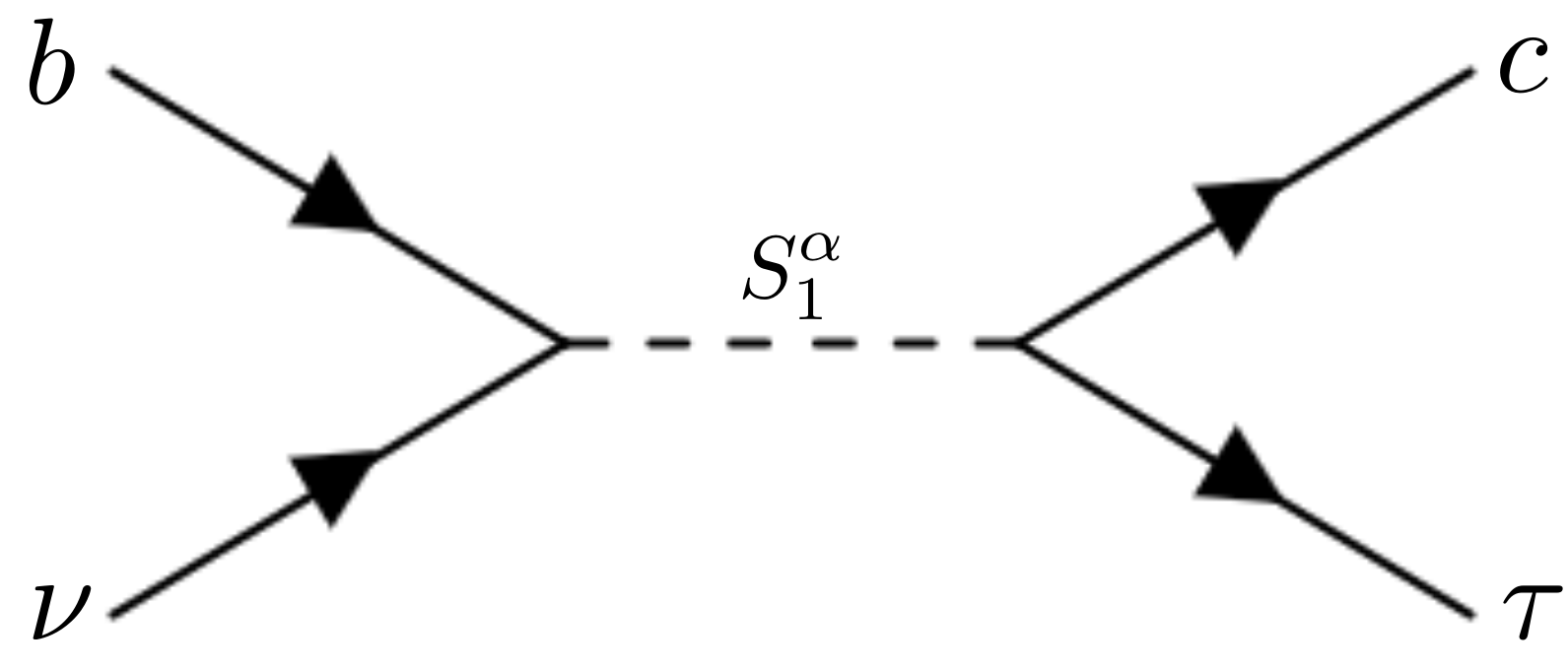}
       \caption{The Feynman diagram contributing to the $b \rightarrow c \tau \nu^i$ process and the $R_{D^{(*)}}$ Anomaly.}
    \label{Fig:RDanomaly}
\end{figure}

\begin{figure}[t!]
\centering
\includegraphics[height=4.5cm, width=6.5cm]{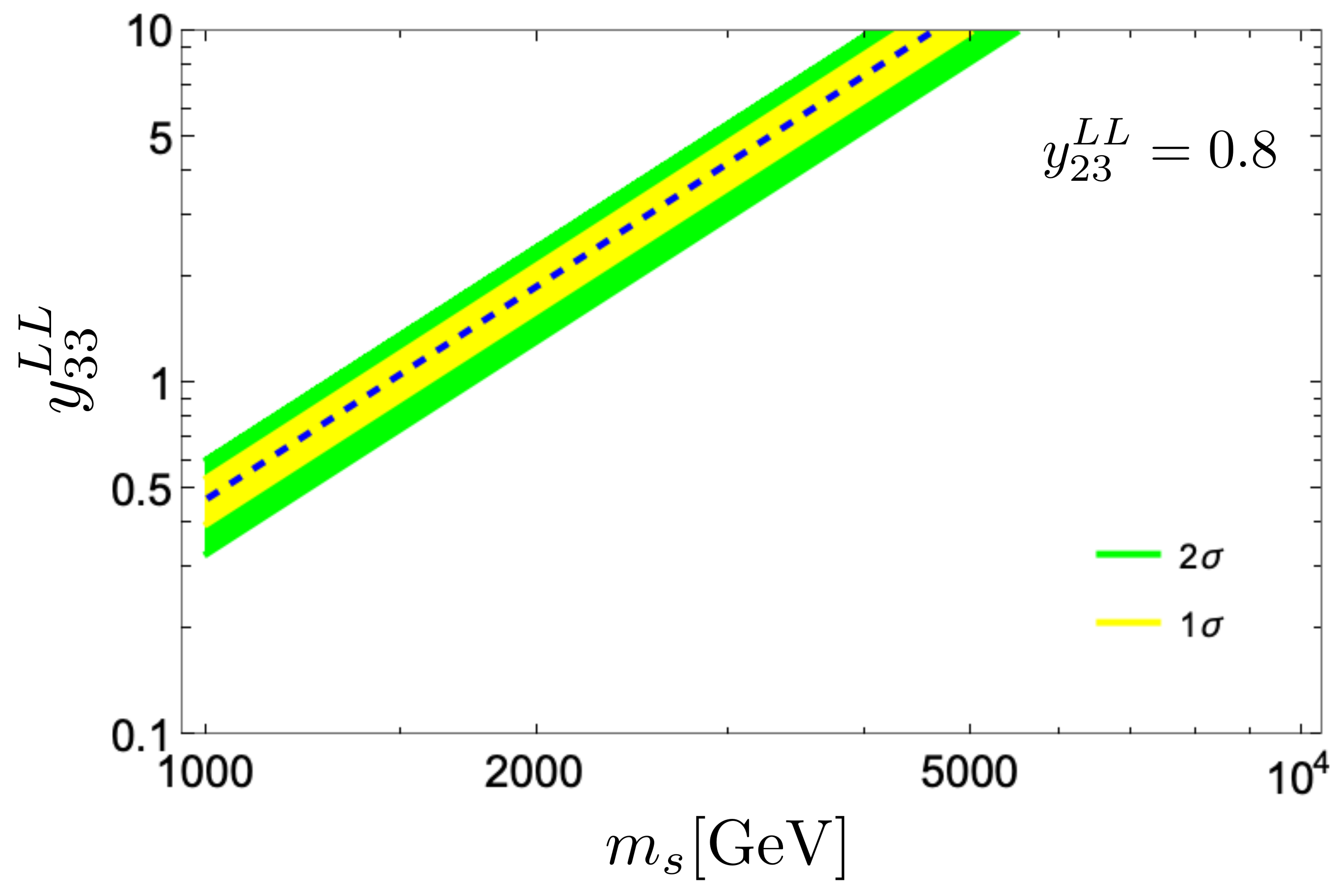}
\caption{The parameter space of the model ($y_{33}^{LL}$ coupling as a function of LQ's mass)  explains $R_{D^{(*)}}$ anomaly. The yellow and green regions indicate the 1$\sigma$ and 2$\sigma$ levels, respectively. The $y_{23}^{LL}$ value is fixed at 0.8. }
\label{Fig:RDlimits}
\end{figure}

The LQs are good candidates in order to explain this anomaly. 
In literature, the different effects of LQs on the B-meson anomalies 
have been studied extensively~\cite{Choi:2018stw,Cheung:2001ip,
Crivellin:2017zlb,Hiller:2017bzc,Buttazzo:2017ixm,Bauer:2015knc,Chen:2017hir,Kumar:2018kmr,Crivellin:2019dwb,Crivellin:2022mff,Cai:2017wry}.
From the Lagrangian in Eq.~\ref{Eq:lagLQ}, the terms relevant to the $R_{D^{(*)}}$ anomaly are,
\beq
\cL \supset -\left(y^{L L} U\right)_{i j} \bar{d}_L^{C i}  S^{\alpha}_{1} \nu_L^j
+\left(V^T y^{L L}\right)_{i j} \bar{u}_L^{C i}  S^{\alpha}_{1} e_L^j + \text{h.c.},
\eeq
that the $ S^{\alpha}_{1}$ LQ can contribute to the $b \rightarrow c \tau \nu^i$ process. 
The relevant Feynman diagram is shown in Fig.~\ref{Fig:RDanomaly}.
According to the economical flavor ansatz (Eq.~\ref{ansatz}) and after integrating out the scalar LQ,
the effective Lagrangian relevant for $b \rightarrow c \tau \nu^i$ is given by~\cite{Dorsner:2016wpm},
\beq
\cL^{\rm{eff}}_{b \rightarrow c \tau \nu^i} =  -\frac{4 G_F}{\sqrt{2}} C^{^{\rm{BSM}}}_{cb}
[(\bar{c}_{L} \gamma^{\mu} b_{L})(\bar{\tau}_{L}\gamma_{\mu} \nu^{\tau}_{L})]+ \text{h.c.},
\eeq
where $C^{^{\rm{BSM}}}_{cb}= \frac{ v^2 (y^{L L}_{33}U_{33})(V^{T}_{22} y^{L L}_{23})^*}{4~m_{_{S^{\alpha}_{1}}}^2} $ 
that $v$ is the SM Higgs vacuum expectation value.
Fig.~\ref{Fig:RDlimits} illustrates the parameter space of the model ($y_{33}^{LL}$ coupling as a function of LQ's mass) where can explain the $R_{D^{(*)}}$ anomaly. According to our benchmark, the value of the $y_{23}^{LL}$ is fixed at 0.8.
This region of parameter space is consistent with the current LHC bound on the LQ mass
from $S^{\alpha}_{1} \rightarrow  \bar{c} \bar{\tau}$ 
and $S^{\alpha}_{1} \rightarrow \bar{b} \nu_\tau$ decay channels~\cite{CMS:2017kil}.

\section{Summary} \label{sec:summary}                                       
In this paper, we introduce a new portal between the standard model (SM) and the dark sectors by scalar leptoquarks (LQ)
to resolve some long-standing anomalies simultaneously.
The SM predicts the branching ratio of the neutron decay to proton, electron, and anti-electron-neutrino is $100\%$,
however, there is an anomaly in the neutron decay width measurements. 
In the bottle experiments, where the number of the remaining neutrons is counted, 
the measured neutron lifetime is shorter than the one measured in beam experiments, 
where the number of the produced protons is counted. 
This anomaly can be solved, if the neutron decays to invisible (for example, particles in the dark sector) with a branching ratio around $1\%$. 
We suggest that the neutron decays into a dark scalar ($\phi$) and an SM anti-neutrino by these scalar LQ mediators.
The dark scalar is singlet under the SM gauge symmetries but it carries the baryon and lepton numbers 
since there are severe constraints on the baryon and lepton numbers violation processes.
The mass of the $\phi$ should be in the narrow range between 937.9 and 939.565 to satisfy all the current bounds.

The $\phi$ with the aforementioned properties can be a good dark matter (DM) candidate. 
We showed that the freeze-in mechanism can produce the dark scalar in the early universe 
and its relic abundance is compatible with the DM relic density measured by the Planck collaboration.
Furthermore, we discussed that this model in good parameter space region can explain other SM observational anomalies simultaneously.
For instance, the anomalous magnetic moment of the muon, and the $R_{D^{(*)}}$ anomaly
can be explained through our model at the same time. 
\section*{Acknowledgments} \label{sec:ack}                                            
We gratefully thank Fatemeh Elahi for the fruitful discussion and her comments that greatly improved the manuscript. 
Also, we are thankful to the CERN theory division for their hospitality.
\bibliographystyle{JHEP}                                                                                      
\bibliography{NeutronAnomaly}

\end{document}